\documentclass[prd,longbibliography,nofootinbib,twocolumn,preprintnumbers]{revtex4-1}

\usepackage[T1]{fontenc}
\usepackage[utf8]{inputenc}
\usepackage{relsize}
\usepackage{soul}
\usepackage{calligra}
\usepackage{verbatim}
\usepackage{caption}
\usepackage{comment}
\usepackage{adjustbox}
\usepackage[titles]{tocloft}
\cftsetindents{section}{0em}{2.5em}
\cftsetindents{subsection}{2.5em}{2.5em}

\usepackage{appendix}

\usepackage{graphicx,epsfig}
\usepackage[usenames,dvipsnames,table]{xcolor}
\usepackage{subfig}
\usepackage{float}
\usepackage{placeins}

\usepackage{tabularx}
\usepackage{makecell,multirow}
\usepackage{dcolumn}
\usepackage{enumitem}

\usepackage{amsmath,amsthm,amssymb}
\usepackage{slashed}
\usepackage{mathrsfs}
\usepackage{bm}
\usepackage{leftidx}
\usepackage{commath}

\usepackage{url}
\usepackage{color}
\definecolor{darkblue}{rgb}{0.0,0.0,0.6}
\definecolor{red}{rgb}{0.9, 0,0}
\definecolor{navy}{rgb}{0.05, 0.05,0.8}
\definecolor{linkcolor}{rgb}{0.0,0.5,0.4}

\usepackage[colorlinks]{hyperref}
\hypersetup{
     colorlinks   = true,
     citecolor    = linkcolor,
     linkcolor    = linkcolor,
     urlcolor    = linkcolor
}
\newcommand{\Eq}[1]{Eq.~\ref{#1}} 
\newcommand{\Eqs}[2]{Eqs.~\ref{#1} and \ref{#2}} 
\newcommand{\Sec}[1]{Sec.~\ref{#1}} 
\newcommand{\Secs}[2]{Secs.~\ref{#1} and \ref{#2}} 
\newcommand{\App}[1]{Appendix~\ref{#1}} 
\newcommand{\Fig}[1]{Fig.~\ref{#1}}

\newcommand{\x}{\chi}

\newcommand{\eV}{\text{eV}}

\newcommand{\MeV}{\text{MeV}}
\newcommand{\GeV}{\text{GeV}}

\newcommand{\cm}{\text{cm}}

\newcommand{\s}{\text{s}}

\newcommand{\kpc}{\text{kpc}}

\newcommand{\Gyr}{\text{Gyr}}

\newcommand{\be}{\begin{equation}}
\newcommand{\ee}{\end{equation}}

\newcommand{\grad}{\vec{\nabla}}
\newcommand{\sv}{\langle \sigma v \rangle}
\newcommand{\cms}{\cm^3 \, \s^{-1}}
\newcommand{\p}{\prime}
\newcommand{\Udark}{U(1)^\prime}
\newcommand{\Ap}{A^\prime}
\newcommand{\mAp}{m_{A^\prime}}
\newcommand{\eps}{\epsilon}
\newcommand{\alp}{\alpha^\p}
\newcommand{\Lag}{\mathscr{L}}

\newcommand{\mdm}{m_{_\text{DM}}}
\newcommand{\rhodm}{\rho_{_\text{DM}}}
\newcommand{\tmw}{t_\text{mw}}
\newcommand{\Tmw}{T_\text{mw}}
\newcommand{\JiDM}{J_\text{iDM}}

\newcommand{\DM}{\text{DM}}
\newcommand{\SM}{\text{SM}}
\newcommand{\eff}{\text{eff}}

\newcommand{\Tkin}{T_\text{kd}}
\newcommand{\Tchem}{T_\x^\text{dec}}
\newcommand{\Tfo}{T_\text{fo}}

\newcommand{\Gkd}{\Gamma_\text{kd}}
\newcommand{\Gcd}{\Gamma_\x}

\newcommand{\ashedit}[1]{\textcolor{red}{[\textbf{AB}:} \textcolor{orange}{\textbf{currently}} \textcolor{Green}{\textbf{editing}} \textcolor{purple}{\textbf{here}]} \\}

\begin{document}

\hfill{\small FERMILAB-PUB-21-457-T}

\title{Reviving MeV-GeV Indirect Detection with Inelastic Dark Matter}

\author{Asher Berlin$^{a,b}$}
\email{aberlin@fnal.gov}
\author{Gordan Krnjaic$^{a,c,d}$}
\email{krnjaicg@fnal.gov }
\author{Elena Pinetti$^{a,d}$}
\email{epinetti@fnal.gov}
\affiliation{$^a$Theory Division, Fermi National Accelerator Laboratory}
\affiliation{$^b$Superconducting Quantum Materials and Systems Center (SQMS),
Fermi National Accelerator Laboratory}
\affiliation{$^c$University of Chicago, Department of Astronomy and Astrophysics}
\affiliation{$^d$University of Chicago, Kavli Institute for Cosmological Physics}


\begin{abstract}

Thermal relic dark matter below $\sim 10 \ \text{GeV}$ is excluded by cosmic microwave background data if its annihilation to visible particles is unsuppressed near the epoch of recombination. Usual model-building measures to avoid this bound involve kinematically suppressing the annihilation rate in the low-velocity limit, thereby yielding dim prospects for indirect detection signatures at late times. In this work, we investigate a class of cosmologically-viable sub-GeV thermal relics with late-time annihilation rates that are detectable with existing and proposed telescopes across a wide range of parameter space. We study a representative model of inelastic dark matter featuring a stable state $\chi_1$ and a slightly heavier excited state $\chi_2$ whose abundance is thermally depleted before recombination. Since the kinetic energy of dark matter in the Milky Way is much larger than it is during recombination, $\chi_1 \chi_1 \to \chi_2 \chi_2$ upscattering can efficiently regenerate a cosmologically long-lived Galactic population of $\chi_2$, whose subsequent coannihilations with $\chi_1$ give rise to observable gamma-rays in the $\sim 1 \ \text{MeV} - 100 \ \text{MeV}$ energy range. We find that proposed MeV gamma-ray telescopes, such as e-ASTROGAM, AMEGO, and MAST, would be sensitive to much of the thermal relic parameter space in this class of models and thereby enable both discovery and model discrimination in the event of a signal at accelerator or direct detection experiments.

\end{abstract}


\maketitle

\section{Introduction}
\label{sec:intro}

Light dark matter (DM) has received much attention over the past decade
as new terrestrial experiments have been proposed
to vastly expand laboratory sensitivity to new physics below the electroweak scale -- for reviews, see Refs.~\cite{Krnjaic:2022ozp,Essig:2022dfa}. Amongst the various possibilities, $\MeV-\GeV$ scale DM in the form of a light thermal relic is especially motivated by basic principles of early-Universe thermodynamics and known examples within the Standard Model (SM)~\cite{Berlin:2018bsc}. 

Thermal DM is in equilibrium with the visible sector at early times and later freezes out after becoming non-relativistic, such that number-changing reactions occur less than once per Hubble time. Although such reactions remain out of equilibrium, rare annihilation events can continue to inject energy into the SM plasma. For instance,  DM annihilations into electromagnetically-charged particles can persist between recombination and reionization, yielding observable distortions in the temperature anisotropies of the cosmic microwave background (CMB). The lack of such deviations sharply constrains some of the most predictive and well-motivated freeze-out models~\cite{Padmanabhan:2005es,Slatyer:2015jla}. This limit can be phrased in terms of the thermally-averaged annihilation cross section $\sv$ to visible states, which at the time of recombination is constrained by the Planck satellite at the level of~\cite{Planck:2018vyg} 
\be
\label{eq:cmbintro}
\sv_\text{cmb} \lesssim 3 \times 10^{-26} \  \cms \   \bigg( \frac{\mdm}{10 \ \GeV} \bigg)
~,
\ee
where $\mdm$ is the DM mass. Since standard thermal DM predicts $\sv \sim 10^{-26} \ \cms$ at the time of freeze-out when the DM was quasi-relativistic, models involving annihilations to visible final states with velocity-independent ($s$-wave) cross sections are excluded for masses below $\sim 10 \ \GeV$.

Of course, this cross section needs not be velocity-independent, and the CMB bound is naturally evaded in models where the annihilation rate is suppressed by the small DM velocity $v$ during recombination. For example, in models with $p$-wave annihilation, $\sv \propto v^2$, so that this rate is significantly smaller at later times when DM is highly non-relativistic. However, this approach\footnote{Although larger Galactic annihilation rates can arise in other models of thermal relics involving resonant~\cite{Feng:2017drg,Bernreuther:2020koj,Brahma:2023psr} or forbidden annihilations~\cite{Griest:1990kh,DAgnolo:2015ujb,DAgnolo:2020mpt}, this is typically promising only when the relative mass-splitting between various dark sector states is fixed to a special value, corresponding to a strong tuning between mass scales that are related to independent model parameters.} also severely diminishes the prospects for indirect detection in the Galaxy, since the local annihilation rate is suppressed by the small virial velocity $v \sim 10^{-3}$, resulting in $\sigma v < 10^{-30} \ \cms$ for standard thermal relics. Since such small
 cross sections are beyond the reach of any near-future probe  (see Refs.~\cite{Bartels:2017dpb,Coogan:2021sjs} and references therein), there is 
 strong motivation to identify well-motivated and testable targets for indirect detection in the $\MeV -\GeV$ mass range.

\begin{figure*}[t]
\centering
\hspace{-1cm}
\includegraphics[width=0.9 \textwidth]{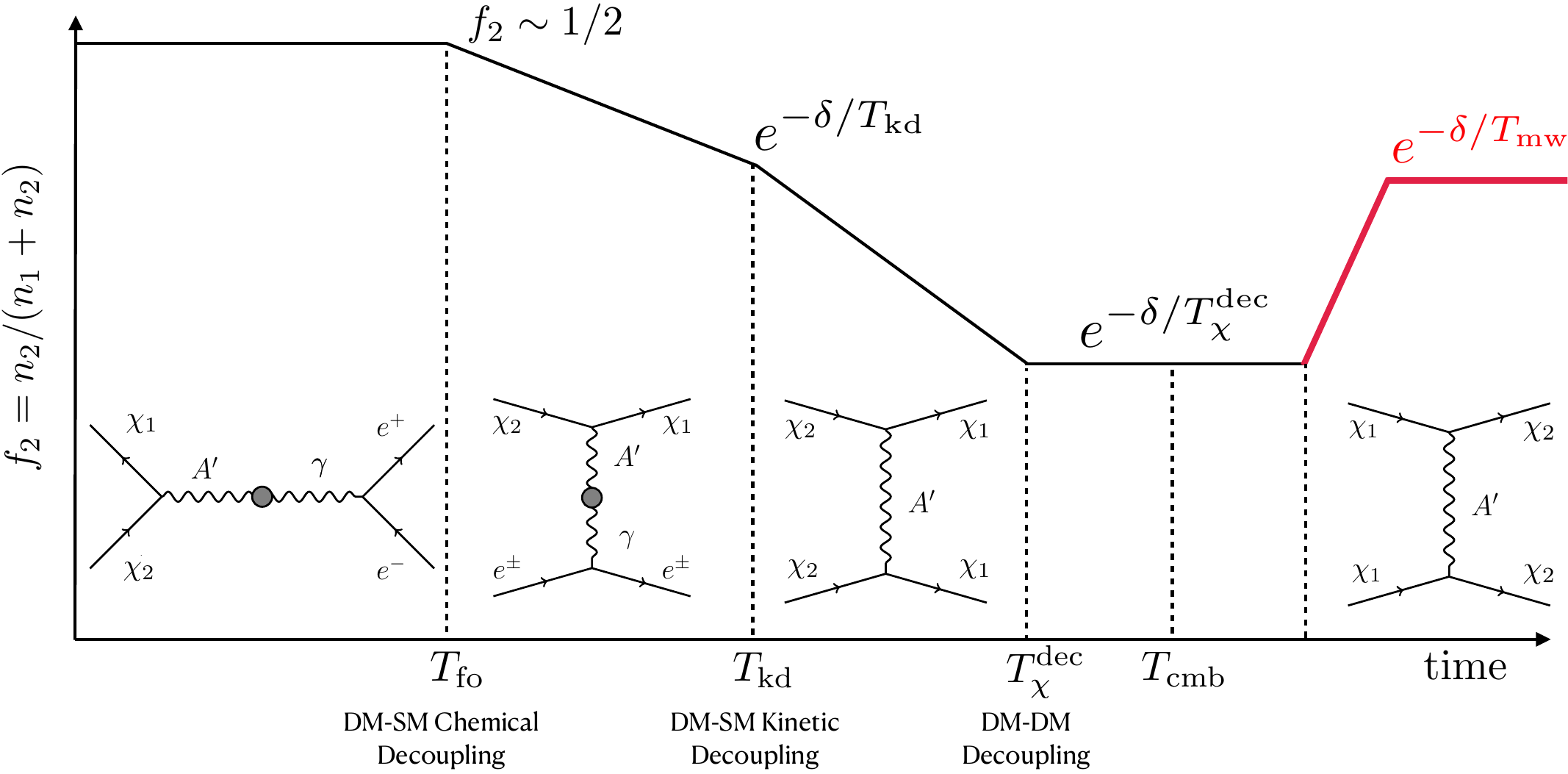}
\vspace{0.2cm}
\caption{A timeline of key events in the cosmic history of iDM. At early times, coannihilations $\x_1 \x_2 \to \SM \ \SM$ freeze out at a temperature much greater than the mass-splitting $\Tfo \gg \delta$, leaving behind roughly equal $\x_{1,2}$ populations with a $\x_2$ number fraction $f_2 = n_2/(n_1+ n_2) \simeq 1/2$. After freeze-out, $\x_2$ continues to be depleted through DM-SM and DM-DM scattering. DM-SM scattering additionally keeps the dark sector in kinetic equilibrium  with the SM until decoupling at a temperature of $\Tkin$. At later times, the heavier state continues to be depleted more rapidly through $\x_2 \x_2 \to \x_1 \x_1$ downscattering until decoupling at a DM temperature of $\Tchem$, such that the primordial fraction of excited states is exponentially suppressed at the time of recombination. As a result, CMB limits on DM annihilations are easily evaded. At much later times, the virialization of DM in the Milky Way increases its velocity, such that $\x_1 \x_1\to \x_2\x_2$ upscattering partially restores the abundance of long-lived heavier states and enhances the coannihilation rate compared to its average cosmological value (red line). This increase implies a local enhancement of $\x_1 \x_2 \to \SM \ \SM$ in the Milky Way whenever the effective Galactic temperature of DM satisfies $\Tmw \gtrsim \Tchem$. If the $\x_2$ population becomes appreciable in the Galaxy, then $\x_1 \x_2 \to \x_1 \x_2$ {\it elastic} scattering can also play an important role in Galaxy evolution.}
\label{fig:timeline}
\end{figure*}

An alternative approach to evading the CMB limit on sub-GeV thermal DM  involves \emph{changing} the DM population between freeze-out and recombination, as in models of ``inelastic DM'' (iDM)~\cite{Tucker-Smith:2001myb}. In this framework, such evolution occurs dynamically in a dark sector containing  a nearly-degenerate pseudo-Dirac DM pair, $\x_1$ and $\x_2$, with a small mass-splitting and an off-diagonal coupling to the SM. The natural size of the mass-splitting is near or below the overall DM mass scale if it arises from the same dynamics responsible for generating the dark sector masses~\cite{Tucker-Smith:2001myb,Arkani-Hamed:2008hhe,Arkani-Hamed:2008kxc}. In this case, DM annihilation proceeds efficiently at early times through coannihilations of $\x_1$ and $\x_2$ to SM final states, but is exponentially suppressed before recombination as the Universe cools and DM self-scattering continues to thermally deplete the abundance of the heavier $\x_2$ state in favor of the lighter $\x_1$ state. A summary of the cosmological history is depicted schematically in \Fig{fig:timeline}.

In this paper, we point out that the kinetic energy of $\x_1$ in the Milky Way is sufficiently large to undo this exponential suppression at late times and thus regenerate a  population of upscattered $\x_2$ excited states that are detectable through their subsequent coannihilation to visible states, $\x_1 \x_2 \to \ \SM \  \SM$. In particular, we highlight the existence of cosmologically-viable and theoretically well-motivated parameter space involving mass-splittings of  $\sim 1 \ \eV - 100 \ \eV$, such that the kinetic energy of $\MeV -\GeV$ scale DM in the Milky Way is sufficient to partially or fully equilibrate a density of cosmologically long-lived excited states. As a result, coannihilations that yield low-energy gamma-rays occur at a rate comparable to (or a few orders of magnitude below) the thermal value for $s$-wave annihilation, which is detectable with existing and proposed indirect detection searches in the Milky Way. In contrast, analogous signals of such processes are suppressed in the early Universe as well as in less dense astrophysical bodies, such as dwarf satellite galaxies, due to the smaller DM velocities involved.

We note that analogous ideas involving the regeneration of late-time indirect detection signals have been previously explored within the context of prompt decays of excited iDM states~\cite{Finkbeiner:2007kk,Chen:2009av,Finkbeiner:2014sja}, late-time decays in a multi-component
hidden sector~\cite{DEramo:2018khz}, and annihilations from asymmetric DM oscillations~\cite{Buckley:2011ye}. While such work is similar in spirit to our study, these models typically involve additional particle content or early-Universe dynamics compared to the model here. By contrast, we present a simple scenario with standard cosmological assumptions where the same interaction that sets the DM thermal relic abundance is also responsible for indirect detection signals in the Galaxy.

The remainder of this paper is organized as follows. In \Secs{sec:overview}{sec:cosmo}, we provide an overview of iDM models and their cosmological history. In \Sec{sec:primordial}, these results are used to determine the primordial fraction of excited states at late times. In \Sec{sec:galaxy}, we study the Galactic dynamics of this scenario and calculate the excited DM population in the Milky Way. These results are  then used in \Secs{sec:IDD}{sec:results} to derive limits from existing indirect detection searches as well as prospects for observing such a population in low-energy gamma-rays with proposed satellites. Finally, we offer concluding remarks in \Sec{sec:conclusion} and discuss directions for future work. A series of appendices, referred to throughout this work, provides additional technical details.

\section{Model Overview}
\label{sec:overview}

In this section, we outline a concrete model of iDM. The DM content of this model involves a Dirac pair of two-component Weyl fermions, $\psi_1$ and $\psi_2$, oppositely charged under a new spontaneously broken $\Udark$ symmetry with gauge interaction 
\be
\label{eq:int_UV}
\Lag \supset e^\p \Ap_\mu \big( \psi_1^\dagger \bar \sigma^\mu \psi_1 -  \psi_2^\dagger \bar \sigma^\mu \psi_2  \big)
~,
\ee
where the dark photon $\Ap$ is the $\Udark$ gauge boson with mass $\mAp$ and gauge coupling $e^\p$. The fermion masses of this theory arise from the Lagrangian terms
\be
\label{eq:Lag1}
\Lag \supset - m_D \, \psi_1 \, \psi_2 - \frac{1}{2} \big( \delta_1 \, \psi_1\psi_1 +   \, \delta_2 \, \psi_2\psi_2 \big) + \text{h.c.}
~,
\ee
where $m_D$ is a gauge-invariant Dirac mass and $\delta_{1,2}$ are $\Udark$ breaking Majorana masses naturally\footnote{Indeed, the Majorana terms are small if they arise from higher-dimensional couplings to the dark Higgs $H^\p$ responsible for spontaneous breaking in the dark sector (e.g., couplings of the form $H^{\p \, 2} \, \psi_{1,2}^2 / \Lambda$ where $\Lambda$ is governed by the mass scale of a heavy singlet that mixes with $H^\p$)~\cite{Tucker-Smith:2001myb,Arkani-Hamed:2008hhe,Arkani-Hamed:2008kxc}. More generally, $\delta_{1,2} \ll m_D$ is technically natural as the theory enjoys an enhanced global symmetry (analogous to SM lepton number) when Majorana masses are absent.} satisfying $\delta_{1,2} \ll m_D$. After diagonalizing this system, the mass eigenstates correspond to a pseudo-Dirac pair,
\be
\x_1 \simeq \frac{i}{\sqrt{2}} \, (\psi_1 - \psi_2)
~~,~~
\x_2 \simeq \frac{1}{\sqrt{2}} \, (\psi_1 + \psi_2)
~,
\ee
with nearly degenerate masses
\be
m_{1,2} \simeq m_D \mp \frac{1}{2} (\delta_1 + \delta_2)
\ee
split by the small amount $\delta \equiv \delta_1 + \delta_2 \ll m_{1,2}$. In terms of  mass eigenstates, the leading interaction of \Eq{eq:int_UV} is
off-diagonal,
\be
\label{eq:Lag2}
\Lag \supset  i e^\p \Ap_\mu \, \bar{\x}_1 \gamma^\mu \x_2 + \mathcal{O} \bigg( \frac{\delta_{1,2}}{m_D} \bigg)
~,
\ee
where we have written $\x_{1,2}$ as four-component Majorana spinors. 

Generically, $\x_1$ and $\x_2$ also couple diagonally to the $\Ap$, but these interactions are suppressed by $\delta / m_1 \ll 1$ and exactly vanish if $\delta_1 = \delta_2$ due to the enhanced charge conjugation symmetry $\x_{1,2} \to\mp \x_{1,2}$, $\Ap \to - \Ap$.
Without loss of essential generality, throughout this paper we assume that $\x_{1,2}$ couple purely off-diagonally to the dark photon, as in \Eq{eq:Lag2}, which governs the dynamics within the dark sector. For instance, \Eq{eq:Lag2} facilitates upscattering $\x_1 \x_1 \to \x_2 \x_2$ as well as \emph{elastic} self-scattering $\x_1 \x_2 \to \x_1 \x_2$ between DM particles, both of which can have significant impacts on the Galactic population, as we will discuss below.

The dark photon coupling to the electromagnetic current arises from a small kinetic mixing between $\Ap$ and the SM photon field $A$,
\be
\label{eq:Lag3}
\Lag \supset \frac{\eps}{2} \, F_{\mu \nu}^\p \, F^{\mu \nu}  + A_\mu J_\text{em}^\mu
~,
\ee
where $F^\p$ and $F$ are the dark and visible field strengths, respectively, $J_\text{em}$ is the SM electromagnetic current, and $\eps \ll 1$ is the dimensionless parameter governing the strength of $\Ap$-$A$ mixing. After rotating away this mixing via $A \to A + \eps \Ap$, SM fields acquire a small coupling to the dark photon through the induced interaction $\eps \, \Ap_\mu J^{\mu}_\text{em}$~\cite{Holdom:1985ag}. 

\section{Cosmological History}
\label{sec:cosmo}

In this section, we review the cosmological history of sub-GeV iDM. The discussion is organized into different subsections, ordered chronologically according to when various processes decouple in the early Universe in the parameter space of interest. In particular, we discuss $\DM-\SM$ chemical decoupling, $\DM-\SM$ kinetic decoupling, and $\x_1 - \x_2$ chemical decoupling, respectively, as depicted in \Fig{fig:timeline}. For previous work investigating such cosmological scenarios, see, e.g., Refs.~\cite{Baryakhtar:2020rwy,CarrilloGonzalez:2021lxm}.

\subsection{$\DM-\SM$ Chemical Decoupling}
\label{sec:freezeout}

Here we discuss DM thermal freeze-out, the process of $\DM-\SM$ chemical decoupling which sets the total $\x_{1,2}$ density in the early Universe. If the kinetic mixing parameter satisfies
\be
\eps \gtrsim 10^{-8} \, \sqrt{\frac{\mAp}{\GeV} }
 ~~, 
\ee
then $\Ap \leftrightarrow \SM \ \SM$ decays and inverse-decays equilibrate the two sectors,  seeding an initially large thermal density of dark sector states~\cite{Evans:2017kti}. 

A priori, there is no preferred ordering of DM and dark photon masses. However, if $\mAp < m_1$ and $e^\p \gg \eps$, then DM freeze-out proceeds predominantly through $\x_1 \x_1 \to \Ap \Ap$ annihilation followed by $\Ap$ decays to visible SM particles.  Since this process is $s$-wave, the CMB limit from \Eq{eq:cmbintro} robustly excludes this mass ordering in the sub-GeV mass range. 

By contrast, if $\mAp > m_1 + m_2$,  then DM annihilations to  on-shell dark photons are kinematically forbidden, and instead the DM relic density is 
governed by the coannihilation $\x_1 \x_2 \to \SM \ \SM$, as shown 
in \Fig{fig:timeline}. In the non-relativistic limit, the cross section for this process is~\cite{Berlin:2014tja,Berlin:2018bsc}
\be 
\label{eq:sigmaFO}
\sv_\text{ann} \simeq   \frac{16 \pi \alpha \alp  \eps^2   m_1^2}{(4 m_1^2 - \mAp^2 )^2}
~,
\ee
where $\alpha$ and $\alp = e^{\p \, 2} / 4 \pi$ are the electromagnetic and  $\Udark$ fine-structure constants, respectively. In our calculations, we incorporate all kinematically accessible channels, including hadronic final states, as described in Refs.~\cite{Izaguirre:2015zva,Berlin:2018jbm}.
Although this cross section is unsuppressed in the low-velocity limit, the CMB bound is alleviated because the $\x_2$ abundance is thermally depleted before recombination for sufficiently large splittings, thereby shutting off $\x_1 \x_2 \to \ \SM \ \SM$ at late times. This is discussed in more detail below.

When the temperature of the early Universe cools to $T = \Tfo \sim m_1 / 10$, the total comoving DM density freezes out to a fixed value. Thus, for $\delta \ll m_1 / 10$, the mass-splitting can be neglected throughout this process, and we follow the calculations of Ref.~\cite{Berlin:2018bsc} to determine the initial relic density of $\x_{1,2}$ particles. In particular, for $\delta \ll m_{1,2} \lesssim 100 \ \MeV$ and $\mAp \gtrsim \text{few} \times m_1$, the $\x_{1,2}$ abundance at freeze-out satisfies   
\be
\label{eq:epsFO}
f_\x \simeq  \bigg( \frac{0.5}{\alp} \bigg) \bigg( \frac{10^{-5}}{\eps} \bigg)^2 \bigg( \frac{m_1}{100 \ \MeV} \bigg)^2 \bigg( \frac{\mAp}{3 \, m_1} \bigg)^4
~,
\ee
where $f_\x$ is the fractional abundance of DM that $\x_1$ and $\x_2$ jointly constitute.

Throughout this work, we focus on the parameter space that satisfies the assumptions leading to \Eq{eq:epsFO}, fixing to the representative values $\mAp / m_1 = 3$ and $\alp = 0.5$, and assuming a standard freeze-out cosmology for $\x_1$ and $\x_2$.\footnote{Thermal freeze-out models with $\mAp / m_1 \lesssim 1$ are typically excluded by measurements of the CMB, as mentioned above, whereas scenarios with $\mAp  / m_1 \gg 3$ or $\alp \ll 1$ are strongly constrained by accelerator searches~\cite{Berlin:2018bsc,Berlin:2020uwy}.} We  also fix the value of $\eps$ by imposing $f_\x = 1$ such that $\x_{1,2}$ account for the entire observed DM abundance, unless mentioned otherwise (in \Sec{sec:results} we will also investigate DM subcomponents, corresponding to $f_\x < 1$). In later sections, we vary the iDM mass-splitting $\delta$ to explore a wide range of possible late-time implications of the $\x_2$ population. However, in the parameter space of interest, it is always the case that $\delta \ll m_1$, so that \Eq{eq:epsFO} is unaffected. 

Furthermore, since we always work in the regime where $\delta < \mAp , 2m_e$, the decays  $\x_2 \to \x_1 \Ap$ and $\x_2 \to \x_1 e^+ e^-$ are kinematically forbidden and $\x_2$ is stable on cosmological timescales;  we neglect the highly subdominant $\x_2 \to \x_1 + 3\gamma$ and $\x_2 \to \x_1  \nu \bar{\nu}$ decay channels, since the corresponding lifetimes are exponentially longer than the age of the Universe~\cite{Batell:2009vb}.

\subsection{$\DM-\SM$ Kinetic Decoupling}
\label{sec:kindec}

After $\DM - \SM$ chemical decoupling, $\DM-\SM$ scattering $\x_1  \, \ell^\pm \leftrightarrow \x_2 \, \ell^\pm$ remains in equilibrium due to the large thermal density of SM leptons,\footnote{Note that there are also similar interactions with quarks and hadrons, but since their densities are exponentially suppressed at temperatures $T \lesssim 100 \ \MeV$, such interactions are subdominant. 
} as shown in \Fig{fig:timeline}. These processes do not change the total DM number, but as long as they are in equilibrium, the DM temperature, $T_\x$, tracks that of the SM, $T$, such that both approximately evolve inversely with the cosmic scale factor $T_\x = T \propto a^{-1}$. 

Since our focus is on sub-GeV DM, the dominant process that maintains kinetic equilibrium after freeze-out is inelastic scattering off of electrons and positrons, whose abundance vastly exceeds that of any other charged species at temperatures $T \lesssim 100 \ \MeV$. The full thermally-averaged thermalization rate for $\x_1  \, e \leftrightarrow \x_2 \, e$ is derived in \App{app:kindec}, but here we briefly summarize its limiting forms. At both high and low temperatures compared to the electron mass, this rate is well approximated by 
\be
\label{eq:kindec1}
\Gkd
\simeq \frac{8  \alpha  \alp  \eps^2 T^2}{3 \mAp^4 m_1}
\times
\begin{cases}
\frac{31}{63} \, \pi^5 \, T^4
& (T \gg m_e)
\\
\frac{16}{\pi} \, m_e^4 \, e^{-m_e / T}
& (T \ll m_e)
~ .
\end{cases}
\ee
The DM and SM sectors kinetically decouple from each other at a temperature of $T_\x = T = \Tkin$ when the typical DM particle no longer efficiently transfers momentum with an electron in a Hubble time. This temperature can be approximated by solving $\Gkd(\Tkin) = H(\Tkin)$, where $H$ is the Hubble expansion rate. After this point, the DM temperature no longer tracks that of the SM plasma, instead redshifting more rapidly as an isolated non-relativistic sector with a temperature scaling of $T_\x \propto a^{-2}$. 

For $m_1 \lesssim 1 \ \GeV$ and our choice of representative model parameters in \Eq{eq:epsFO}, $\Gkd\gg H$ at temperatures comparable to or greater than the electron mass. Thus, $\x_1 e^\pm \leftrightarrow \x_2 e^\pm$ scattering steadily depletes the $\x_2$ population until $T \sim m_e$ when the electron/positron number density becomes Boltzmann suppressed and the scattering rate falls out of equilibrium. Note that for $m_1 \lesssim 10 \ \MeV$, DM annihilations, as discussed in \Sec{sec:freezeout}, decouple after $\DM-\SM$ scattering, so that $\Tkin > \Tfo$ and the two earliest processes in \Fig{fig:timeline} decouple in the opposite order relative to what is shown. In this case, once $\DM$ annihilations to the SM freeze out, the temperatures of the two sectors evolve independently. Generally, we expect this to only hold approximately, and we reserve a more careful numerical treatment, involving solving the relevant Boltzmann equations, to future work.

\subsection{$\DM-\DM$ Chemical Decoupling}
\label{sec:chemdec}

DM self-scattering $\x_1 \x_1 \leftrightarrow \x_2 \x_2$ (as shown in \Fig{fig:timeline}) is unsuppressed by the small coupling $\eps$, and thus can remain in equilibrium long after the processes of the previous two subsections have already decoupled. This implies that even for splittings as small as $\delta \sim \eV$, $\x_2$ can be thermally depleted well before recombination once $T_\x \lesssim \delta$.

As derived in \App{app:chemdec}, the thermally-averaged rate for $\x_2  \, \x_2 \to \x_1 \, \x_1$ scattering is given by
\be
\label{eq:chemdec1}
\Gcd \simeq \frac{8 \pi \, \alpha^{\p \, 2} \, m_1^{3/2}}{\mAp^4} \, n_1 \, e^{-\delta/T_\x} \, \max \bigg( \frac{\delta}{2} \, , \, \frac{T_\x}{\pi}\bigg)^{1/2}
~,
\ee
where $n_1$ is the $\x_1$ number density.\footnote{We note that \Eq{eq:chemdec1} agrees with the expressions in Ref.~\cite{CarrilloGonzalez:2021lxm}, aside from a factor of $\pi$ discrepancy for $T_\x \lesssim \delta$.} We estimate the DM temperature $\Tchem$ at which $\x_1$ and $\x_2$ chemically decouple 
from each other by evaluating $\Gcd (\Tchem) = H$. After decoupling, the primordial abundance of excited states is fixed to
\be
\label{eq:f2}
f_2 = \frac{n_2}{n_1 + n_2} \simeq  \frac{e^{-\delta/\Tchem}}{1 + e^{-\delta/\Tchem}}  
~,
\ee
where we have used detailed balance to relate the number densities of $\x_2$ and $\x_1$ just prior to decoupling, $n_2 \simeq e^{-\delta / T_\x} \, n_1$. 

\subsection{Primordial Excited Fraction}
\label{sec:primordial}

%
\begin{figure}[t]
\centering
\includegraphics[width=0.49 \textwidth]{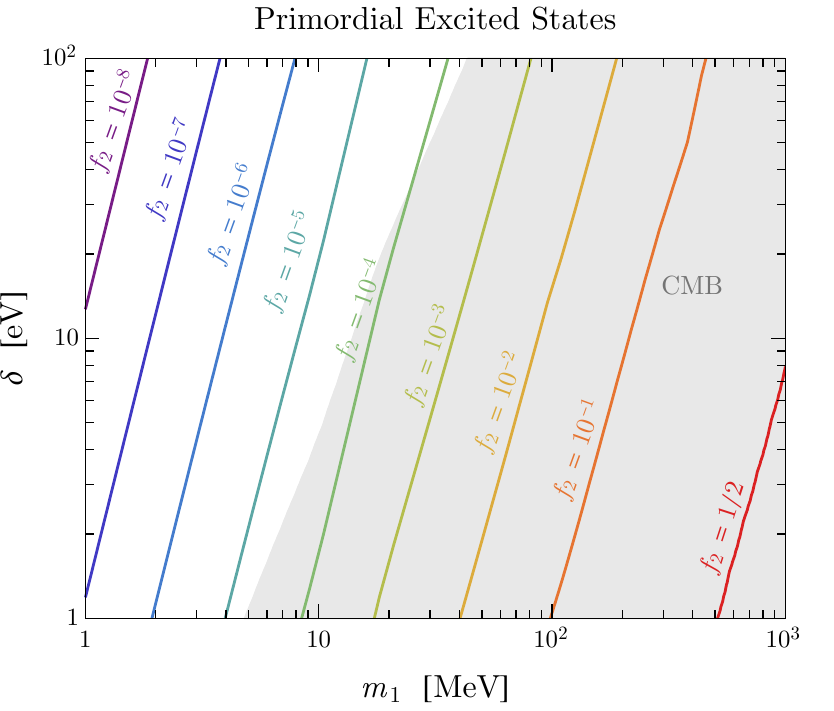}
\caption{Contours of fractional $\x_2$ abundance  $f_2 = n_2/(n_1 + n_2)$ set by $\x_1 - \x_2$ decoupling in the early Universe, as a function of $m_1$ and $\delta$. Here, we have fixed $\mAp / m_1 = 3$, $\alp = 0.5$, and $\eps$ such that the freeze-out abundance of $\x_{1,2}$ agrees with the observed DM density. The gray shaded region corresponds to parameter space excluded by CMB annihilation bounds~\cite{Planck:2018vyg}. }
\label{fig:f2}
\end{figure}

Using the formalism of Secs.~\ref{sec:freezeout}, \ref{sec:kindec}, and \ref{sec:chemdec}, we determine the total $\x_{1,2}$ density, temperature evolution, and $\x_2$ fractional density $f_2$, respectively. In \Fig{fig:f2}, we show $f_2$ as a function of $m_1$ and $\delta$. For fixed $\delta$, heavier dark sector masses suppress the rate $\Gcd$ of \Eq{eq:chemdec1}, such that $\x_1-\x_2$ interconversion decouples at earlier times,  enhancing the relative $\x_2$ density. On the other hand, for larger splittings, the relative abundance of excited states is increasingly Boltzmann suppressed, yielding smaller values of $f_2$. This cosmological value of $f_2$ is fixed at late times, as shown schematically in \Fig{fig:timeline} by the lower plateau  after $T_\x = \Tchem$. 

This primordial $\x_2$ fraction governs the rate for $\x_1 \x_2 \to \SM \ \SM$ near the time of recombination, which is strongly constrained by CMB data. In particular, Planck bounds such coannihilations to $e^+e^-$ final states according to~\cite{Planck:2018vyg}
\be
\label{eq:cmb}
2  f_\x^2  f_1  f_2  \sv_\text{ann} \lesssim 2 \times 10^{-26} \ \cms  \, \bigg(\frac{m_1}{30 \, \GeV}\bigg)
~,
\ee
where the factor of two on the left-hand-side of \Eq{eq:cmb} accounts for the fact that the limit in Ref.~\cite{Planck:2018vyg} assumed annihilation of self-conjugate DM particles, whereas coannihilations of $\x_1$ and $\x_2$ involve distinct species. We have also rescaled the limit by $f_\x^2 f_1 f_2$, where $f_\x$ from \Eq{eq:epsFO} accounts for the DM fraction that $\x_1$ and $\x_2$ jointly constitute, and  $f_1 = 1 - f_2$ is the relative $\x_1$ fraction defined in analogy with $f_2$ in \Eq{eq:f2}. Annihilations to muon, tau, and hadronic final states are also similarly constrained, but are weaker by roughly a factor of $\sim 1/3$ due to their smaller efficiency in transferring energy into the intergalactic medium near the time of recombination~\cite{Slatyer:2015jla}. The resulting constraint from the CMB is shown in \Fig{fig:f2} as the shaded gray region.

\section{Galactic Population}
\label{sec:galaxy}

After dark sector scattering decouples at $\Tchem$, the average cosmological fraction $f_2$ of the heavier $\x_2$ state remains fixed for all subsequent times as the DM temperature continues to fall as $a^{-2}$ due to Hubble expansion. However, in overdense regions where galactic structure forms, the local DM temperature can vastly exceed its average cosmological value due to virialization. Thus, in galaxies and clusters where the local temperature satisfies $T_\x \gtrsim \delta$, there is an additional late-time population of excited states generated by $\x_1 \x_1 \to \x_2 \x_2$ upscattering, which can exceed the cosmological fraction $f_2$ computed in the previous section. In this section, we discuss how such a population can arise in the Milky Way.

\subsection{Galactic Model}
\label{sec:density}

Since the Galactic $\x_2$ fraction is exponentially sensitive to the DM velocity, we begin by describing a toy model for the Milky Way. We assume that upon galaxy formation, the $\x_2$ population has the same cosmological relic fraction $f_2$ set by dynamics in the early Universe (see \Sec{sec:cosmo}) and starts off with the same type of Galactic density profile as the $\x_1$ population. 

To simplify our treatment, we approximate the phase-space of DM in the Milky Way as following that of a Maxwell-Boltzmann distribution, with an \emph{effective} temperature $\Tmw$ that is a function of the galactocentric radius $r$. In particular, using the virial theorem, we take 
\be
\Tmw (r) = \frac{G  M_\text{enc}(r)}{3 r}  \, m_1   
~,
\ee
where $M_\text{enc}(r)$ is the total mass enclosed within $r$. $M_\text{enc}$ is determined directly from our assumed DM density profile as well as the baryonic mass density $\rho_B$. We approximate the stellar bulge, thin disk, and thick disk contributions to $\rho_B$ from the best-fit spherically-symmetric Hernquist profile model advocated for in Refs.~\cite{McMillan:2011wd,Kaplinghat:2013xca},  
\be
\rho_B(r) = \frac{\rho_{B0} \, r_0^4}{ r(r+r_0)^3}
~,
\ee
where $\rho_{B0} = 26 \ \GeV \, \cm^{-3}$ and $r_0 = 2.7 \ \kpc$. 

Later in \Sec{sec:IDD}, we will investigate the impact of various  choices for the DM density profile. However, for concreteness, in this section we model the initial $\x_1$ mass density with a Navarro-Frenk-White (NFW) profile~\cite{Navarro:1996gj}
\be
\rho_1(r)  = \frac{\rho_s}{(r/r_s) \, (1+ r/r_s)^2}
~,
\ee
where $r_s = 20 \ \kpc$ is the scale radius and $\rho_s$ is normalized to yield a local DM density of $0.4 \ \GeV \, \cm^{-3}$ at the position of the Solar System, $r = r_\odot  \simeq 8.5 \ \kpc$.

\subsection{Galactic Evolution}
\label{sec:gal-boltz}

The time-evolution of the Galactic $\x_2$  population is governed by the Boltzmann equation 
\be
\label{eq:GenBoltz}
\frac{\partial {\cal F}_2}{\partial t} + \vec{v} \cdot \grad {\cal F}_2 + \vec{g} \cdot \frac{\partial  {\cal F}_2}{\partial \vec{v}} = C [ {\cal F}_{1,2}]
~,
\ee
where ${\cal F}_{1,2}$ are the phase space densities for $\x_{1,2}$, $\vec{v}$ is the $\x_2$  velocity, $\vec{g}$ is the Galactic gravitational acceleration field, and the collision term $C[{\cal F}_{1,2}]$ accounts for all $\x_{1,2}$ scattering processes. Since solving \Eq{eq:GenBoltz} requires detailed numerical simulations, we  simplify the analysis by factorizing the processes of \emph{inelastic} $\x_2$ production and gravitational evolution, examining upscattering with a simple toy model in this section. Modifications to the $\x_{1,2}$ distributions due to gravitational dynamics and elastic scattering will be discussed later in \Sec{sec:halocomp}, where we consider a wide range of possible scenarios.

\begin{figure}[t]
\centering
\includegraphics[width=0.49 \textwidth]{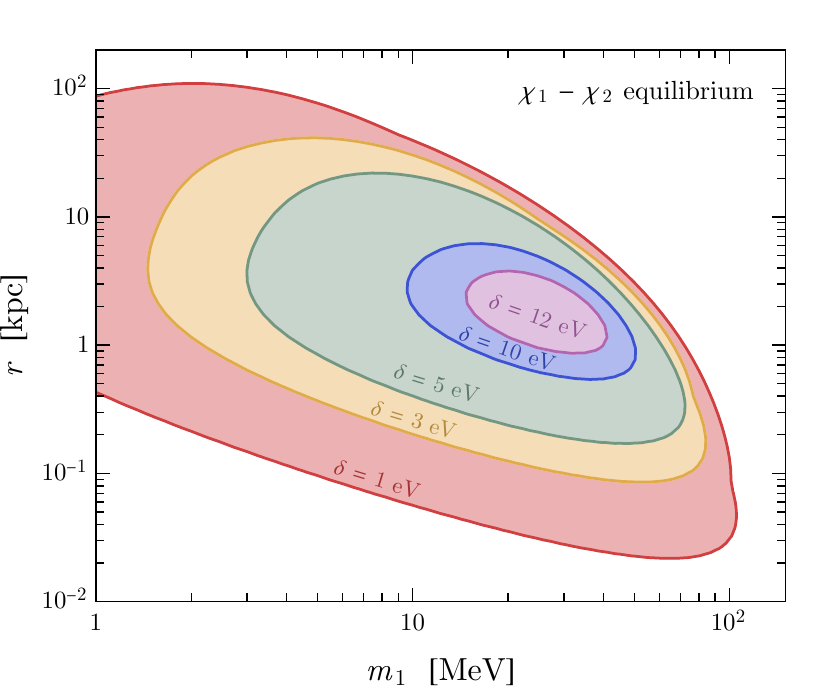}
\caption{Radial regions in the Milky Way in which inelastic dark scattering $\x_1 \x_1 \leftrightarrow \x_2 \x_2$ reenters equilibrium over the age of the Galaxy $\tmw \simeq 13.5 \ \Gyr$, as a function of $m_1$ and for various representative choices of $\delta$, fixing $\mAp / m_1 = 3$ and $\alp = 0.5$. We have also assumed an initial NFW profile for $\x_1$. For each choice of $m_1$ and $\delta$, there is a shell of galactocentric radii within which equilibration occurs.}
\label{fig:requil}
\end{figure}

We approximate $\x_2$ production by taking the zeroth moment of \Eq{eq:GenBoltz} and neglecting gradient and gravitational terms, which amounts to solving
\be
\label{eq:n2dot}
\frac{\partial n_2}{\partial t} =   n_1^2 \, \sv_{1 \to 2} -  n_2^2 \, \sv_{2 \to 1}
~,
\ee
where $n_{1,2}$ is the Galactic number density of $\x_{1,2}$. Here, the thermally-averaged cross section for  $\x_1 \x_1 \to \x_2 \x_2$ upscattering is 
\be
\label{eq:1to2}
\sv_{1 \to 2} = e^{-2 \delta/\Tmw} \, \sv_{2 \to 1}~,
\ee
where $\sv_{2 \to 1}$ is the thermally-averaged cross section for 
$\x_2 \x_2 \to \x_1 \x_1$ downscattering. As derived in \App{app:chemdec}, $\sv_{2 \to 1}$ is approximately
\be
\label{eq:2to1}
\sv_{2 \to 1} \simeq \frac{8 \pi \, \alpha^{\p \, 2} \, m_1^{3/2}}{\mAp^4} \,  \, \max \bigg( \frac{\delta}{2} \, , \, \frac{\Tmw}{\pi}\bigg)^{1/2}
~,
\ee

From \Eq{eq:n2dot}, $n_2$ asymptotes to a constant value  once it evolves to $n_2 \simeq e^{- \delta/\Tmw} \, n_1$, as required by detailed balance. In the limit that the initial $\x_2$ density is small, we can therefore approximate the solution of \Eq{eq:n2dot} as
\be
\label{eq:n2evolve}
n_2 \simeq n_1 \, \min \big( N_\text{scatt} \, , \, e^{-\delta/\Tmw} \big)
~,
\ee
where $N_\text{scatt}  = n_1 \, \sv_{1 \to 2} \, t$ is the number of upscatters per $\x_1$ particle after time $t$. Note that $\x_2$ approaches detailed balance once the average number of upscatters is $N_\text{scatt} \sim e^{-\delta / \Tmw}$, which occurs on a characteristic timescale of 
\be
t \sim \frac{e^{\delta / \Tmw}}{n_1 \, \sv_{2 \to 1}}
~.
\ee
Due to its strong temperature dependence, the upscattering cross section $\sv_{1 \to 2}$ is peaked around a characteristic radial range in the Galaxy, driven by the fact that $\Tmw$ peaks at intermediate radii. In \Fig{fig:requil}, we show the radial regions of the Milky Way in which $\x_2$ reenters chemical equilibrium with $\x_1$ over the age of the Galaxy, $\tmw \simeq 13.5 \ \Gyr$, as a function of $m_1$ and for various choices of the splitting $\delta$. We see that for mass-splittings $\delta \lesssim 10 \ \eV$, $\x_2$ is able to reenter chemical equilibrium over a significant portion of the Galaxy.

\begin{figure}[t]
\centering
\includegraphics[width=0.49 \textwidth]{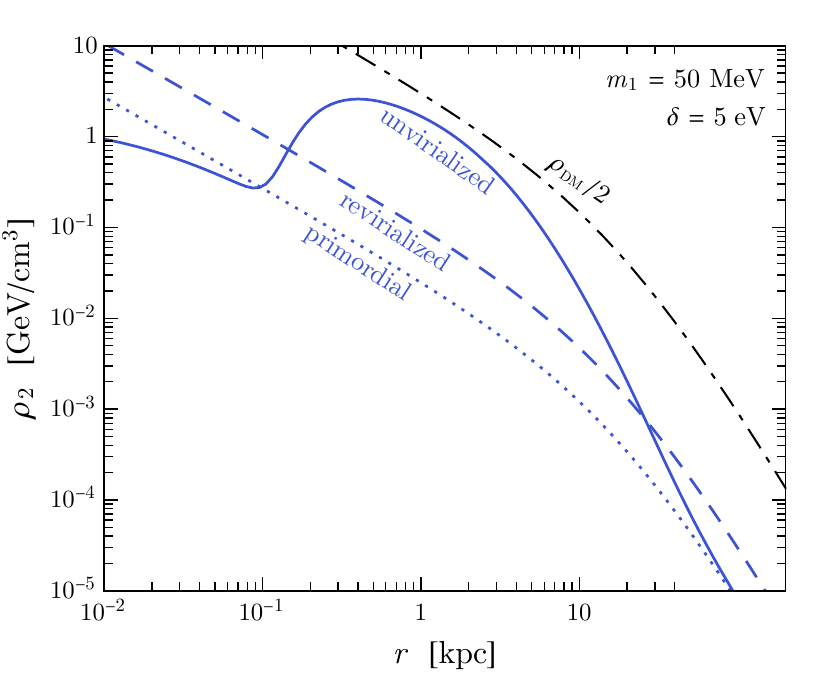}
\caption{Radial mass density profiles of $\x_2$ in the Milky Way, for $m_1 = 50 \ \MeV$, $\delta = 5 \ \eV$, $\mAp / m_1 = 3$, $\alp = 0.5$, and assuming an initial NFW profile for $\x_1$ and $\x_2$. For the solid line labelled ``unvirialized,'' we assume that $\x_2$ remains at the same radius as the progenitor $\x_1$ particle. For the dashed line labelled ``revirialized,'' we instead assume that the $\x_2$ population readjusts to an NFW profile with normalization fixed to conserve particle number.  For reference, we also show as a dotted line the initial primordial $\x_2$ profile (see \Eq{eq:ninitial}) and as a dot-dashed line a standard NFW DM density profile, scaled down by a factor of two.
}
\label{fig:rho2}
\end{figure}

We now calculate the Galactic density profile of $\x_1$ and $\x_2$ at $t = \tmw$ by numerically solving \Eq{eq:n2dot}, along with $n_1 + n_2 = \text{constant}$, at each radius $r$, starting from the initial condition at $t = 0$ where 
\be
\label{eq:ninitial}
n_{1,2} (0) = f_{1,2} \, f_\x \, \rhodm / m_{1,2}
~,
\ee
for a given Galactic DM profile $\rhodm$. Note that this treatment neglects the impact of additional effects, including  elastic $\x_1 \x_2 \to \x_1 \x_2$ scattering and gravitational dynamics, which will be addressed later in \Sec{sec:halocomp}. In \Fig{fig:rho2}, we show the resulting Galactic mass density profile of $\x_2$, $\rho_2 = m_2 \, n_2$, fixing $m_1 = 50 \ \MeV$ and $\delta = 5 \ \eV$  and for two different assumptions regarding the Galactic evolution of excited states. In particular, for the solid line in \Fig{fig:rho2}, we take $\x_2$ to be produced along orbits of fixed radius. By contrast, for the dashed line we instead assume that after production, the entire Galactic $\x_2$ population readjusts to an NFW profile with normalization fixed to conserve the total $\x_2$ number. These two scenarios are referred to as ``unvirialized'' and ``virialized,'' respectively.\footnote{Note that our use of ``virialization'' is simply meant to imply the process of gravitational relaxation/evolution.} For reference, we also show as a dotted line the initial primordial $\x_2$ profile (see \Eq{eq:ninitial}) and as a dot-dashed line a standard NFW DM density profile (scaled down by a factor of two). The peak-like feature near $r \sim 1 \ \kpc$ shows the growth to the $\x_2$ population generated from $\x_1 \x_1 \to \x_2 \x_2$ Galactic upscattering at late times. Note that for much larger radii, the distribution does not significantly evolve from its original primordial value since the low density of $\x_1$ particles suppresses the likelihood to upscatter over the age of the Galaxy. Instead, for much smaller radii, the small Galactic temperature favors downscattering compared to upscattering, reducing the density of excited states in that region.

%

\section{Indirect Detection}
\label{sec:IDD}

The last section discussed the Galactic density profiles for $\x_1$ and $\x_2$. Here, we use these results to determine the resulting flux of visible particles from DM coannihilations $\x_1 \x_2 \to \SM \ \SM$. We begin by providing some general formalism in \Sec{sec:genform}. In \Sec{sec:halocomp},  we then investigate the sensitivity of our final results to the assumed form of the $\x_1$ and $\x_2$ density profiles.

\subsection{General Formalism}
\label{sec:genform}
 
In conventional indirect detection studies, the flux $\Phi_\gamma$ of photons from DM annihilations is determined by
\be
\label{eq:flux}
\frac{d\Phi_\gamma}{dE_\gamma} = c_\x \, \frac{\sv_\text{ann}}{4 \pi \, \mdm^2} \, \frac{dN_\gamma}{dE_\gamma} \, J
~,
\ee
where $\sv_\text{ann}$ is the thermally-averaged  annihilation cross section in the Galaxy, $dN_\gamma/dE_\gamma$ is the number of photons per annihilation event per unit energy $E_\gamma$, and the $J$-factor depends on the Galactic mass density of DM, $\rhodm$. The constant $c_\x$ incorporates additional factors arising from the counting of DM species. For example, $c_\x = 1/2$ for Majorana DM in order to not overcount distinct pairs of self-conjugate particles, and $c_\x = 1/4$ for Dirac DM, since annihilations involve interactions between distinct particles and antiparticles, each of which account for half of the total DM density. For standard DM annihilations, the $J$-factor is given by
\be
\label{eq:JDM}
J = \int d \Omega \, d \ell ~ \rhodm(r)^2
~,
\ee
where $\Omega$ and $\ell$ are solid angle and radial coordinates with respect to a local observer, which are related to the galactocentric radial coordinate by 
\be
r^2 = r_\odot^2 + \ell^2 - 2 \, r_\odot \, \ell \cos{\theta}
~.
\ee

In our iDM scenario, there are two distinct DM species that coannihilate to yield a flux of SM particles, similar to Dirac DM. However, unlike the Dirac case, the $\x_2$ population that arises from Galactic upscattering can have a  distinct profile, such that we define the $J$-factor for iDM as
\be
\label{eq:JiDM}
\JiDM \equiv \int d \Omega \, d \ell ~ \rho_1 (r) \, \rho_2 (r)
~,
\ee
where the mass density profiles for $\x_1$ and $\x_2$ are determined from \Sec{sec:gal-boltz}. Note that in the limit where $m_1 \simeq m_2$ and $\rho_1 (r) \simeq \rho_2 (r) \simeq \rhodm / 2$, $\JiDM \simeq J / 4$. With this convention for $\JiDM$, $c_\x = 1$ for iDM in \Eq{eq:flux}.

Most indirect detection studies of the Milky Way estimate the reach of various existing or future telescopes by presenting the sensitivity to the annihilation cross section $\sv$, assuming an NFW profile of self-conjugate DM particles. To make direct contact with these studies, we define an \emph{effective} iDM annihilation cross section,
\be
\label{eq:sveff}
\sv_\eff \equiv  2 \, \sv_\text{ann} ~ \bigg( \frac{\JiDM}{J_\text{NFW}} \bigg)
~,
\ee
where $J_\text{NFW}$ is defined as in \Eq{eq:JDM} for an NFW DM profile. The numerical prefactor in \Eq{eq:sveff} is fixed so that a telescope that is sensitive to self-conjugate (e.g., Majorana) DM annihilations with an NFW density profile and an annihilation cross section of $\sv_\eff$ is also sensitive to iDM particles of the same mass and annihilation channels.

\subsection{$J$-factor Profile-Dependence}
\label{sec:halocomp}

%
\begin{figure}[t]
\centering
\includegraphics[width=0.49 \textwidth]{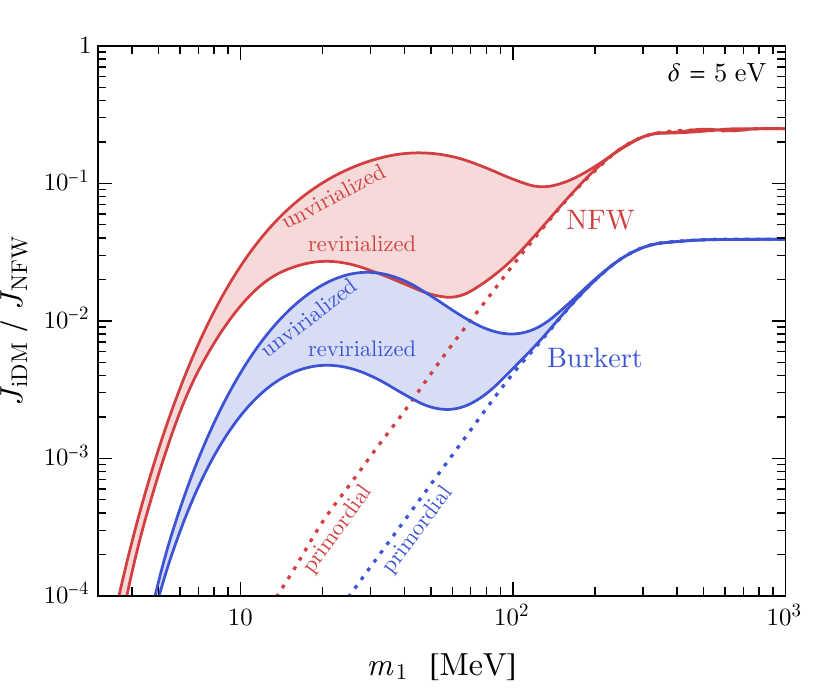}
\caption{As solid lines, the Milky Way $J$-factor for iDM (see \Eq{eq:JiDM}) normalized by that of conventional NFW-distributed DM (see below \Eq{eq:sveff}), as a function of $m_1$, fixing $\delta = 5 \ \eV$, $\mAp / m_1 = 3$, and $\alp = 0.5$. We have computed the $J$-factor by integrating over the polar angles $0^\circ \leq \theta \leq 20^\circ$.  $\JiDM / J_\text{NFW} = 1/4$ indicates that iDM coannihilations yield the same flux as NFW-distributed Dirac DM annihilating with the same cross section. For the red or blue lines, we assume an initial $\x_{1,2}$ distribution described by an NFW profile or a $10 \ \kpc$ core Burkert profile, respectively. The upper boundary of each shaded band labeled ``unvirialized'' gives the $J$-factor assuming $\x_2$ particles maintain the same radius as their $\x_1$ progenitors, whereas for the lower boundary of each band labeled ``revirialized'' we assume that the $\x_2$ population eventually settles into the same type of Galactic distribution as $\x_1$ with normalization fixed to conserve particle number. Also shown as dotted lines are the contributions from the initial primordial density of excited states, i.e., taking $\rho_{1,2}$ according to their initial values in \Eq{eq:ninitial}.}
\label{fig:JiDM}
\end{figure}

Self-scattering processes can lead to significant modifications to the $\x_{1,2}$ density profiles. For instance, if the $\x_2$ population is sufficiently abundant, elastic scattering $\x_1 \x_2 \to \x_1 \x_2$ can result in a  cored central density, as typically found in models of self-interacting DM (SIDM)~\cite{Tulin:2017ara}. However, the novel upscattering process $\x_1 \x_1 \to \x_2 \x_2$ alters the $\x_2$ kinematics to counteract such core formation, since it increases the likelihood for such particles to cluster in the inner Galaxy. This is discussed briefly in \App{app:upscatterhalo} and confirmed in the recent $N$-body simulation of Ref.~\cite{ONeil:2022szc}, which used model parameters comparable to the smallest DM masses and mass-splittings that we consider in this work. Fully modeling all of these effects requires a dedicated $N$-body simulation in our parameter space of interest, which is beyond the scope of this work. Thus, in this section we approximate the impact  of these processes on the iDM $J$-factor by considering a wide range of possible density profiles.

\begin{figure}[t]
\centering
\includegraphics[width=0.49 \textwidth]{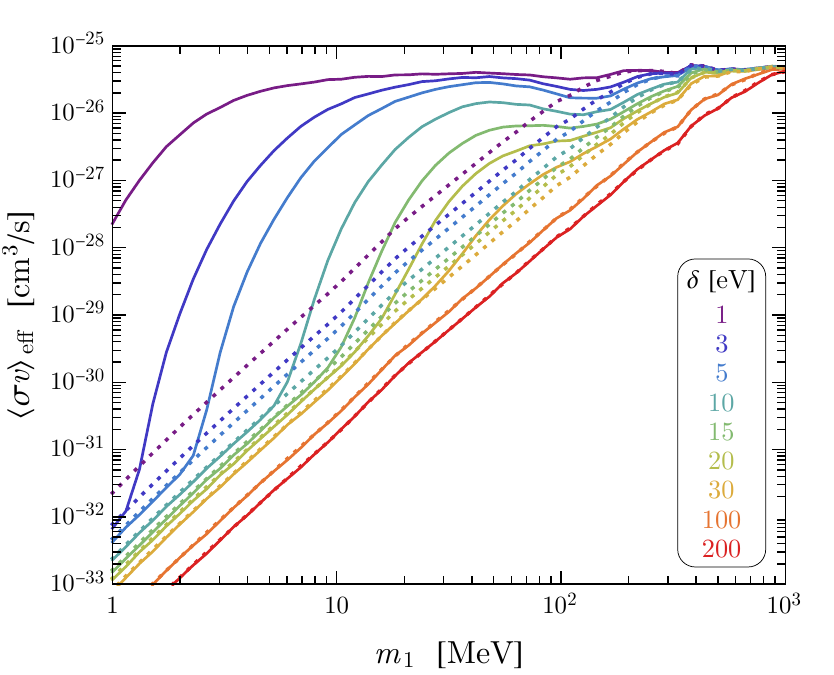}
\caption{The effective annihilation cross section $\sv_\eff$ (see \Eq{eq:sveff}), as a function of $m_1$ and for various choices of $\delta$, fixing $\mAp / m_1 = 3$, $\alp = 0.5$, and $\eps$ such that the freeze-out abundance of $\x_{1,2}$ agrees with the observed DM density. The dotted lines show only the contribution from the cosmological relic population of $\x_2$ parameterized by $f_2$ in \Eq{eq:f2}, whereas the solid lines also include the additional contribution from Galactic upscattering in the Milky Way (assuming an initial NFW profile for $\x_1$ and $\x_2$ and that the upscattered $\x_2$ particles stay fixed to the radii at which they are produced).}
\label{fig:sveff}
\end{figure}
\begin{figure}[t]
\centering
\includegraphics[width=0.47 \textwidth]{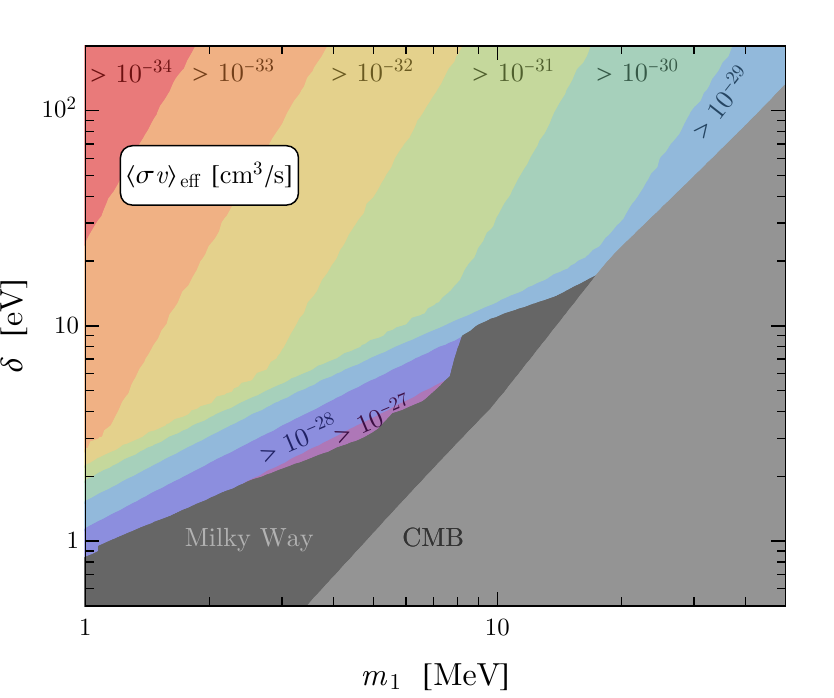}
\caption{Contours of the total effective cross section $\sv_\eff$ (with contributions from both the primordial and Galactic upscattered $\x_2$ populations), as a function of $m_1$ and $\delta$, with the same model assumptions as in \Fig{fig:sveff}. Also shown in gray are regions excluded by observations of the CMB~\cite{Planck:2018vyg} and indirect detection searches of the Milky Way~\cite{Essig:2013goa,Boudaud:2016mos,Cirelli:2020bpc, Cirelli:2023tnx}.}
\label{fig:mass_delta}
\end{figure}

To determine $\JiDM$, we use $\rho_1$ and $\rho_2$ as calculated in \Sec{sec:galaxy}. \Fig{fig:JiDM} shows the ratio $\JiDM / J_\text{NFW}$ as a function of $m_1$ for $\delta = 5 \ \eV$. Here, we compute the $J$-factor after integrating over a range of polar angles corresponding to $0^\circ \leq \theta \leq 20^\circ$ and assuming either an initial NFW or Burkert $\x_1$ profile. In the latter case, the overall flux of the signal is suppressed by our choice of the largest core size typically considered for a Burkert profile, corresponding to $\sim 10 \ \kpc$~\cite{Nesti:2013uwa} (this is meant to illustrate the maximal effect that coring from self-scattering can have on $\JiDM$). Under each of these profile assumptions, we investigate two further possibilities, corresponding to either a $\x_2$ population that tracks the same radius at which it is produced or instead one that readjusts to the same type as the original $\x_1$ profile, fixing the normalization to conserve total $\x_2$ number (see the discussion in \Sec{sec:galaxy}). These two choices bracket the range of $\JiDM$ values shown in \Fig{fig:JiDM}. Also shown as dotted lines are the contributions to the $J$-factor solely from the initial primordial density of excited states, i.e., taking $\rho_{1,2}$ according to their initial values in \Eq{eq:ninitial}. Thus, we see that uncertainties regarding the effects of gravitational  dynamics can significantly impact the estimated flux of annihilation products, at the level of a couple orders of magnitude within our parameter space of interest.

In the remainder of this study, we therefore adopt the more optimistic assumptions, corresponding to an initial NFW profile for $\x_1$ and a population of upscattered $\x_2$ excited states that tracks the radial position of their $\x_1$ progenitors. 
However, we note that neither of the bands shown in \Fig{fig:JiDM}
take into account that the upscattered $\x_2$ particles generically have less energy\footnote{Although the total energy is conserved in these reactions, the $\x_2$ produced from upscattering have less combined kinetic and potential energy than their $\x_1$ progenitors.} than their progenitors, which may lead to greater $\x_2$ densities near the Galactic Center and thereby enhance the $J$-factor relative to what is shown here.  We expand upon this phenomenon briefly in \App{app:upscatterhalo} and leave a more detailed study to future work.

\begin{figure*}[t]
\centering
\includegraphics[width=0.49 \textwidth]{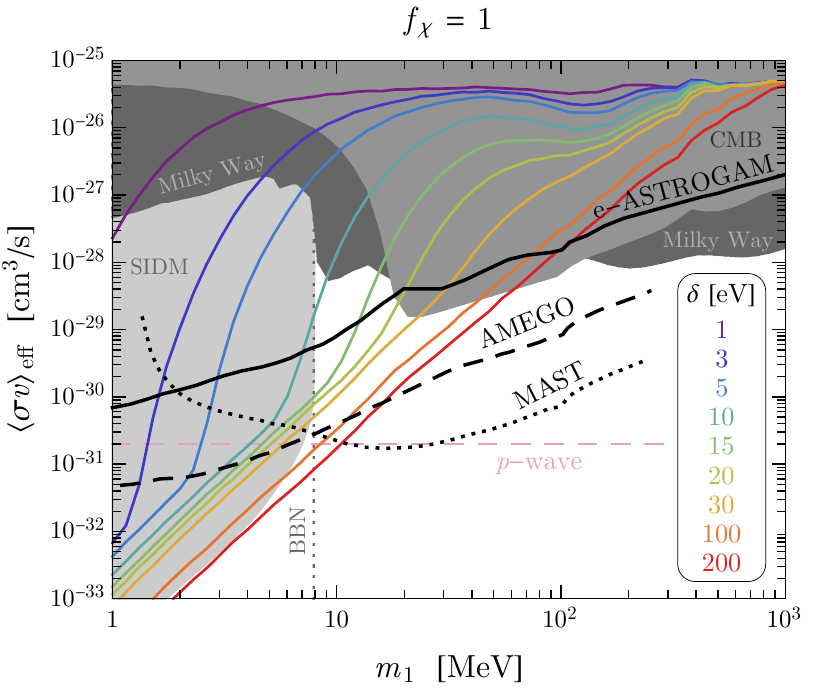}
\includegraphics[width=0.49 \textwidth]{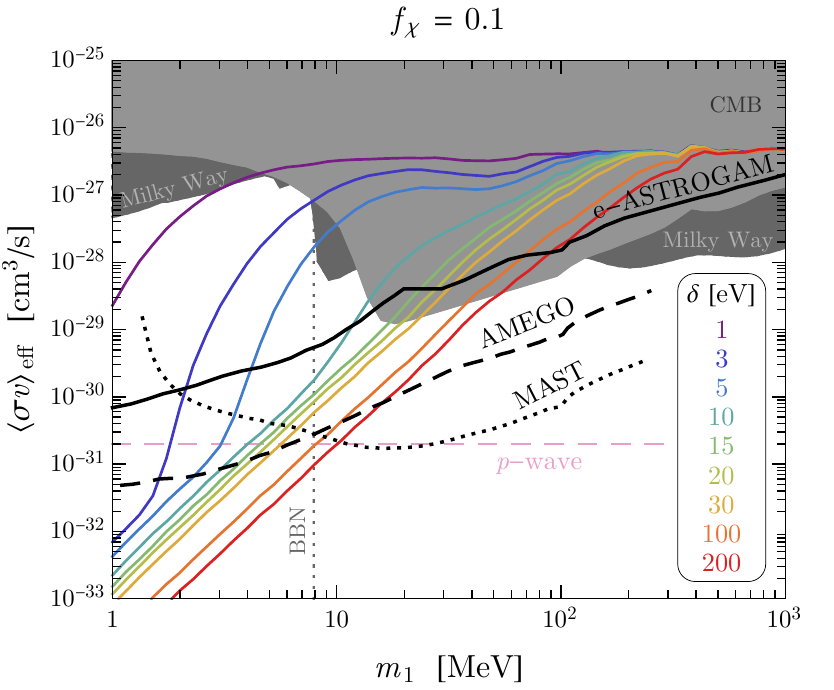}
\caption{ {\bf Left}: The effective cross section $\sv_\eff$ (as in \Eq{eq:sveff} and \Fig{fig:sveff}), as a function of $m_1$ for various choices of $\delta$, fixing $\mAp / m_1 = 3$, $\alp = 0.5$, and $\eps$ such that the freeze-out abundance of $\x_{1,2}$ agrees with the observed DM density ($f_\x  = 1$). The initial distributions for $\x_1$ and $\x_2$ are taken to be NFW profiles, and we assume that the upscattered $\x_2$ stay fixed to the radii at which they are produced. Three representative projections of proposed gamma-ray satellites are shown as solid, dashed, and dotted black lines, including e-ASTROGAM~\cite{e-ASTROGAM:2017pxr}, AMEGO~\cite{Caputo:2022xpx}, and MAST~\cite{Dzhatdoev:2019kay}, respectively.
Also shown as shaded gray regions are constraints from the CMB~\cite{Planck:2018vyg}, indirect detection searches of the Milky Way~\cite{Essig:2013goa,Boudaud:2016mos,Cirelli:2020bpc,Cirelli:2023tnx}, and DM self-interactions in galaxy clusters (labeled ``SIDM''). The lower bound on $m_1$, as derived from the successful predictions of BBN under the assumption of a standard cosmology~\cite{Giovanetti:2021izc}, is shown as a vertical gray dotted line. The approximate rate for $p$-wave annihilating thermal DM is shown as a dashed pink line. Direct detection limits from iDM downscattering constrain splittings larger than those considered here, $\delta \gtrsim 200 \ \eV$~\cite{Baryakhtar:2020rwy,CarrilloGonzalez:2021lxm}. {\bf Right}: Same as the left panel, but instead fixing $\eps$ such that the freeze-out abundance of $\x_{1,2}$ constitutes only 10\% of the total DM abundance ($f_\x  = 0.1$). In this case, constraints on self-interacting DM do not apply.}
\label{fig:reach}
\end{figure*}
%

\section{Results}
\label{sec:results}

In \Fig{fig:sveff}, we show $\sv_\eff$ (see \Eq{eq:sveff}) as a function of $m_1$ and for various choices of $\delta$; the dotted lines show the contribution from only the initial primordial population of excited states. For sufficiently large splittings, upscattering in the Galaxy is exponentially rare, such that the primordial population dominates the flux of coannihilations at late times. From \Eq{eq:1to2}, for a fixed value of $\delta$, upscattering is exponentially suppressed at low masses and power-law suppressed at high masses, resulting in an optimal range of DM masses for the Galactic contribution to $\sv_\eff$. On the other hand, for the primordial population, $\sv_\eff$ grows with increasing mass due to the enhancement in $f_2$ (see the discussion in \Sec{sec:primordial}). This is evident in \Fig{fig:sveff}, which shows that for splittings $\delta \lesssim 30 \ \eV$ and intermediate masses $m_1 \sim 1 \ \MeV - 100 \ \MeV$, coannihilations arising from the Galactic upscattered $\x_2$ particles dominantly contribute to $\sv_\eff$. Note that for the primordial component, $f_2 \simeq 1/2$ for $m_1 \gtrsim 100 \ \MeV$, such that late-time annihilations are unsuppressed and $\sv_\eff$ asymptotes to the standard freeze-out value of $\sv_\eff \simeq 5 \times 10^{-26} \ \cms$ for sub-GeV masses~\cite{Steigman:2012nb}.

\Fig{fig:mass_delta} displays $\sv_\text{eff}$ in the two-dimensional parameter space spanned by $m_1$ and $\delta$. In light gray we show regions of parameter space excluded by observations of the CMB (as discussed in \Sec{sec:primordial}). Also shown in dark gray are constraints derived from existing observations of the Milky Way. In particular, these consist of limits from gamma-ray observations by INTEGRAL and COMPTEL~\cite{Essig:2013goa}, cosmic-ray measurements by the Voyager 1 spacecraft~\cite{Boudaud:2016mos}, and X-ray satellites~\cite{Cirelli:2020bpc,Cirelli:2023tnx}. This region of parameter space corresponds to scenarios in which $\x_2$ efficiently reenters equilibrium in the Galaxy, thus sourcing an annihilation rate that is comparable to typical $s$-wave annihilations of thermal DM. 

Proposed gamma-ray satellites will be able to significantly extend this sensitivity to larger $\delta$, corresponding to smaller $\sv_\eff$. This is evident in \Fig{fig:reach}, which shows as colored lines the predicted sizes of $\sv_\eff$ as a function of mass, for various choices of $\delta$. In the left- and right-panels, we fix $f_\x = 1$ and $f_\x = 0.1$, respectively. Also shown as a dashed pink line is the approximate rate for $p$-wave thermal DM assuming $f_\x = 1$. Three representative projections of proposed gamma-ray satellites are shown as solid, dashed, and dotted black lines, including e-ASTROGAM~\cite{e-ASTROGAM:2017pxr}, AMEGO~\cite{Caputo:2022xpx}, and MAST~\cite{Dzhatdoev:2019kay}, respectively. In particular, we have taken the e-ASTROGAM projections from Ref.~\cite{Bartels:2017dpb} and the AMEGO and MAST projections from Ref.~\cite{Coogan:2021sjs}. We note that  Ref.~\cite{Coogan:2021sjs} adopted an Einasto DM profile; we have rescaled their projections appropriately to account for the smaller $J$-factor of an NFW profile. Such instruments have the capability to probe a wide range of currently unexplored thermal iDM parameter space, with annihilation rates orders of magnitude larger than standard $p$-wave annihilating DM. Although not shown in \Fig{fig:reach}, the GECCO~\cite{Orlando:2021get} and COSI~\cite{Zoglauer:2021coa} instruments may also have competitive sensitivity in this parameter space. 

In addition to bounds from the CMB and existing indirect detection searches of the Milky Way, for $f_\x = 1$ we also display limits from considerations of self-interacting DM (SIDM). In particular, the large relative DM velocity $v \sim 10^{-2}$ in observed galaxy clusters, such as the Bullet cluster~\cite{Robertson:2016xjh}, implies that DM as light as $m_1 \sim 1 \ \MeV$ has enough kinetic energy to upscatter $\x_1 \x_1 \to \x_2 \x_2$ for splittings as large as $\delta \sim 100 \ \eV$. The region labeled ``SIDM'' in the left panel of \Fig{fig:reach} corresponds to parameter space where upscattering occurs at a rate of $ \sigma_{1 \to 2} / m_1 \gtrsim 10 \ \cm^2 / \text{g}$. In this case, such processes could lead to large deviations in the  density profiles of merging galaxy clusters. However, note that since typical studies of SIDM consider elastic scattering, qualitative differences could arise for mass-splittings comparable to the DM kinetic energy, such that properly recasting this bound in the iDM parameter space necessitates a more careful treatment. Moreover, since observations limit the total mass lost in such systems to be less than $\sim 10 \%$, iDM self-interactions are completely unconstrained if they make up less than $\sim 10 \%$ of the DM~\cite{Markevitch:2003at}. This scenario is considered in the right-panel of \Fig{fig:reach}, where we instead consider a DM subcomponent with $f_\x = 0.1$. Also note that in this case, the asymptotic behavior of $\sv_\eff$ at high masses following from \Eqs{eq:epsFO}{eq:sveff} is given by $\sv_\eff \propto  f_\x^2 \, \sv_\text{ann} \propto f_\x$.

For $m_1 \lesssim 10 \ \MeV$, the thermal density of dark sector particles can alter the expansion rate near temperatures of $T \sim 1 \ \MeV$, leading to predictions for the light element abundances that are in conflict with standard Big Bang nucleosynthesis (BBN). In particular, assuming a standard cosmological history, measurements of the CMB and inferred primordial nuclear abundances exclude $m_1 \lesssim 8 \ \MeV$ for the dark sectors considered in this work~\cite{Giovanetti:2021izc}. This constraint is shown as a dotted gray line in \Fig{fig:reach}, but can be ameliorated if additional new physics modifies the photon-neutrino temperature ratio or introduces additional relativistic species at early times. We leave a more complete model investigation along these lines to future work.

We conclude this section with a brief discussion on the sensitivity of direct detection searches for DM scattering. In this work, we have refrained from considering mass-splittings larger than $\delta = 200 \ \eV$, since for  $200 \ \eV \lesssim \delta \lesssim 2 m_e$, downscattering of primordial excited states off of electrons, $\x_2 e \to \x_1 e$, is highly constrained by existing XENON1T, SuperCDMS, and CRESST data~\cite{Baryakhtar:2020rwy,CarrilloGonzalez:2021lxm}. Instead, for $\delta \lesssim 10^{-6} \, m_1$, upscattering in the form of $\x_1 e \to \x_2 e$ can be searched for in low-threshold targets, but the limited exposure of current experiments is insufficient to probe the thermal parameter space investigated here~\cite{Berlin:2018bsc}. Provided that low-backgrounds can be maintained with larger exposures, future projections for the OSCURA~\cite{oscura} and SuperCDMS~\cite{supercdmscollaboration2023strategy} experiments could provide sensitivity to such parameter space.

\section{Discussion}
\label{sec:conclusion}

We have explored a class of cosmologically-viable DM models in which the same interaction that fixes the relic abundance also generates MeV-GeV gamma-rays in the Galaxy that are detectable across a wide range of parameter space. In this scenario, visible DM annihilations proceed efficiently in the early Universe through the coannihilation between a lighter $\x_1$ and heavier $\x_2$ state, but are exponentially suppressed near the time of recombination as DM self-scattering continues to thermally deplete the $\x_2$ density. At much later times, the increased kinetic energy of DM particles in the Milky Way enables $\x_1 \x_1 \to \x_2 \x_2$ upscattering to efficiently regenerate a population of long-lived excited states. Their resulting coannihilations give rise to low-energy gamma-rays that are within reach of future telescopes, such as e-ASTROGAM~\cite{e-ASTROGAM:2017pxr}, AMEGO~\cite{Caputo:2022xpx}, and MAST~\cite{Dzhatdoev:2019kay}. A novel feature of this scenario is that  upscattering, followed by coannihilation, is exponentially-dependent on the DM temperature, which peaks at intermediate radii in the Milky Way. As a result, for certain DM mass-splittings, there exists a radial shell within which $\x_1$ and $\x_2$ reenter chemical equilibrium over the age of the Galaxy.

Notably, such theories of inelastic DM serve as an important model-target for accelerator and direct detection searches~\cite{Krnjaic:2022ozp}. Hence, indirect detection may play a key role in enabling model discrimination if signals of sub-GeV DM arise (or fail to arise) elsewhere. For instance, observing a signal in gamma-rays and in low-energy accelerator experiments,  such as LDMX~\cite{LDMX:2018cma,Berlin:2017ftj} or Belle-II~\cite{Campajola:2021pdl}, would provide evidence in favor of this scenario against other models involving, e.g., $p$-wave annihilations (as is the case for Majorana fermions or complex scalars annihilating to the SM through an intermediate dark photon~\cite{Berlin:2017ftj}).

In regards to direct detection searches, future experiments such as OSCURA~\cite{oscura} and SuperCDMS~\cite{supercdmscollaboration2023strategy} may ultimately be sensitive to such models, either from downscattering $\x_2 e \to \x_1 e$ or upscattering $\x_1 e \to \x_2 e$, depending on the mass-splitting.\footnote{Note that direct detection projections for inelastic (i.e., pseudo-Dirac) DM  often assume that the leading signal arises from loop-induced elastic scattering, which is beyond the reach of upcoming experiments. However, as shown in Refs.~\cite{Baryakhtar:2020rwy,CarrilloGonzalez:2021lxm}, downscattering can lead to larger signals, depending on the local density of $\x_2$. 
} 
Since in such models there is a unique prediction for the relative strength of scattering signals and the Galactic annihilation flux, mutually consistent signals from all three processes would constitute a smoking gun discovery for this scenario and allow the determination of the underlying model parameters.

In addition to what we have investigated in this work, there remain open questions regarding cosmological and astrophysical predictions of these models. Below, we highlight some future directions warranting further study along these lines.
\\

\noindent \textbf{Galactic Dynamics}:
As noted above, our treatment of $\x_1 \x_1 \to \x_2 \x_2$ Galactic upscattering is based on an analytical model of the Milky Way, which 
makes several idealized assumptions. To quantify the theoretical uncertainty associated with our modeling, we have considered a wide range of possibilities for the $\x_2$ distribution in the Galaxy (see \Fig{fig:JiDM}). Nevertheless, there are several important effects that cannot be easily estimated using our approach. For example, we separately consider scenarios in which the upscattered $\x_2$ population either redistributes into the same halo profile as the $\x_1$ population or, alternatively, tracks the same radius where it was originally produced. However, both possibilities ignore the fact that the $\x_2$ are less energetic than the $\x_1$, due to upscattering. A brief discussion of this point is given in \App{app:upscatterhalo}. A proper treatment of this effect is beyond the scope of this paper and calls for a dedicated $N$-body numerical simulation, along the lines of Ref.~\cite{ONeil:2022szc} but geared more towards the specific parameter space investigated in our work. Since including additional $\x_2$ clustering would likely increase the Galactic $J$-factor, the analytical projections in this paper are conservative.
\\

\noindent \textbf{Cosmological Evolution}:
Our calculation of the primordial $\x_2$ fraction adopts the instantaneous decoupling approximation in which $\x_2$ number-changing processes ($\x_1 e \to \x_2 e$ and $\x_1 \x_1 \to \x_2 \x_2$)  maintain chemical equilibrium until their rates fall below the Hubble scale. While this suffices for an order of magnitude estimate of the cosmological $\x_2$ number density, a more accurate treatment requires numerically solving a system of Boltzmann equations for $\x_{1,2}$. We leave such an analysis to future work.
\\

\noindent \textbf{Cluster \& Dwarf Comparison}: 
For a given mass-splitting, a minimum amount of kinetic energy is required to inelastically upscatter, and thus such processes are highly sensitive to the effective DM temperature. As a result, the types of signals discussed here may vary drastically across astrophysical systems, such as dwarf galaxies and galaxy clusters, whose velocity dispersions differ considerably from that of the Milky Way. In particular, for the parameter space investigated in this work, such signals are exponentially-suppressed in dwarf galaxies due to the smaller DM velocities involved~\cite{McConnachie:2012vd,ONeil:2022szc}. Conversely, it would be especially interesting to study the sensitivity of proposed low-energy gamma-ray telescopes to signals originating from galaxy clusters~\cite{Lisanti:2017qlb}, as their sizeable velocity dispersions~\cite{Damsted:2023kem} could provide access to models involving much larger mass-splittings. We defer a full treatment of these issues to future work.

\bigskip


\section*{Acknowledgements}
We would like to thank Saniya Heeba, Dan Hooper, Katelin Schutz, and Oren Slone for useful conversations. Fermilab is operated by the Fermi Research Alliance, LLC under Contract DE-AC02-07CH11359 with the U.S. Department of Energy. This material is based upon work supported by the U.S. Department of Energy, Office of Science, National Quantum Information Science Research Centers, Superconducting Quantum Materials and Systems Center (SQMS) under contract number DE-AC02-07CH11359 and by the Kavli Institute for Cosmological Physics at the University of Chicago through an endowment from the Kavli Foundation and its founder Fred Kavli. This work was completed in part at the Perimeter Institute. Research at Perimeter Institute is supported in part by the Government of Canada through the Department of Innovation, Science and Economic Development Canada and by the Province of Ontario through the Ministry of Colleges and Universities.

\newpage
\appendix
\onecolumngrid

\section{$\DM - \SM$ Kinetic Decoupling}
\label{app:kindec}

Here, we provide discussion regarding $\DM - \SM$ scattering and kinetic decoupling.  In the limit that $\mAp \gg \sqrt{s} \gg \delta$ and $m_1 \gg m_e , T$, the differential cross section for $\x_2 e \to \x_1 e$ is given by
\be
\frac{d \sigma_{\x e}}{dt} \simeq \frac{\pi \alpha \alp \eps^2}{2 \mAp^4 m_1^2 \, p_e^2} \, \Big[ 2 (s - m_1^2 - m_e^2)^2 + 2 s t + t^2 \Big]
~,
\ee
where $s$ and $t$ are the usual Mandelstam variables and $p_e$ is the electron momentum. Following Refs.~\cite{Gondolo:2012vh,Bertoni:2014mva}, the rate for maintaining kinetic equilibrium is then given by
\be
\Gkd \simeq - \frac{1}{3 m_1 T} \, \int \frac{d^3 p_e}{(2 \pi)^3} ~ f_e (1-f_e) \, v_e \, \int_{-4 p_e^2}^0 \hspace{-0.3 cm} dt ~ t \, \frac{d \sigma_{\x e}}{dt}
~,
\ee
where $f_e = (e^{E_e/T} + 1)^{-1}$ is the phase-space distribution of electrons with energy $E_e$,  and $v_e$ is the electron velocity. In the limit $T \gg m_e$ or $T \ll m_e$, the above expression reduces to \Eq{eq:kindec1}. Note that schematically, the above rate is $\Gkd \sim n_e \sigma_{\x e} v_e   / N_\text{scatt}$, where $N_\text{scatt} \sim m_1 T / q^2 \gg 1$ is the number of scatters, each exchanging momentum $q$, needed to transfer a total amount comparable to the DM momentum $\sim \sqrt{m_1 T}$.

\section{$\DM - \DM$ Chemical Decoupling}
\label{app:chemdec}

Here, we provide more discussion regarding DM downscattering $\x_2 \x_2 \to \x_1 \x_1$. To begin, we calculate the cross section in the limit that $\mAp \gg \sqrt{s}$ to be
\be
\sigma_{2 \to 1} \simeq \frac{2 \pi \alpha^{\p \, 2}}{3 s \mAp^4} \, \sqrt{\frac{s - 4 m_1^2}{s - 4 m_2^2}} ~ \Big[ 7 s^2 - 4 s \, (4 m_1^2 + 4 m_2^2 + 3 m_1 m_2) + 76 m_1^2 m_2^2 \Big]
~.
\ee
To evaluate the thermal average $\sv_{2 \to 1}$, we use Eq.~3.8 of Ref.~\cite{Gondolo:1990dk}. In the limit that $\delta \ll T_\x \ll m_1$ or $T_\x \ll \delta \ll m_1$, we arrive at \Eq{eq:2to1}. The corresponding downscattering rate of \Eq{eq:chemdec1} is then given by $\Gcd = n_2 \, \sv_{2 \to 1}$, where $n_2 = e^{-\delta / T_\x} \, n_1$.

\section{Upscattering in the Milky Way Halo}
\label{app:upscatterhalo}

Since upscattering $\x_1 \x_1 \to \x_2 \x_2$ converts kinetic energy into mass energy, it causes particles to fall more deeply into gravitational wells, enhancing the central density of halos, analogous to the dynamics associated with dissipative reactions. This was noted in the detailed numerical simulations of Ref.~\cite{ONeil:2022szc}, where it was found that enhanced central densities from upscattering can counteract coring from general scattering-induced heat transport throughout the halo. Here, we provide a brief semi-analytic discussion of the central density enhancement from upscattering, leaving a more dedicated analysis to future work. 

We denote the total energy of an initial $\x_1$ particle as $E = m_1 + T + U$, where $T$ and $U$ are its kinetic and gravitational potential energy, respectively. For convenience, let us rewrite this as $E = m_1 + \tilde{E}$, where $\tilde{E} \equiv T + U$. Using the virial theorem, $T = - \frac{1}{2} U$, we can rewrite $\tilde{E}$ solely in terms of the gravitational energy at radius $r$, 
\be
\label{eq:Etilde1}
\tilde{E} = \frac{1}{2} \, U = - \frac{G M_\text{enc}(r) \, m_1}{2 r}
~,
\ee
where $M_\text{enc}(r)$ is the total Galactic mass enclosed within $r$. Using that the total energy $E$ is conserved as $\x_1$ upscatters and converts into the slightly heavier state $\x_2$, we can relate the change in $\tilde{E}$ to the change in mass $m_\x$ of the particle, 
\be
\label{eq:Etilde2}
\frac{d \tilde{E}}{dt} = - \frac{d m_\x}{dt} = - \Gamma_{1 \to 2} ~ \delta
~,
\ee
where $\Gamma_{1 \to 2} = n_1 \, \sv_{1 \to 2}$ is the upscattering rate per $\x_1$ particle. Alternatively, the change in $\tilde{E}$ can be rewritten using the chain rule as
\be
\label{eq:Etilde3}
\frac{d \tilde{E}}{dt} = \frac{d \tilde{E}}{dr} \, \frac{dr}{dt} = \frac{G M_\text{enc}(r) \, m_1}{2 r^2} \, \frac{dr}{dt}
~,
\ee
where we used \Eq{eq:Etilde1} in the last equality. Equating \Eqs{eq:Etilde2}{eq:Etilde3}, we have that the change in radius from upscattering is given by
\be
\label{eq:drdt}
\frac{dr}{dt} \simeq - 2 \, \Gamma_{1 \to 2} ~ \frac{\delta}{m_1} \, \frac{r^2}{G M_\text{enc}(r)}
~.
\ee

Given a general profile for the enclosed mass $M_\text{enc} (r)$ and an initial radius $r_i = r(0)$ at $t = 0$, \Eq{eq:drdt} can be solved for the radius today $r(\tmw)$. This can then be inverted to obtain $r_i (r)$. Since upscattering does not change total particle number, we then relate the initial number density profile at early times $n_i (r_i)$ to the profile today $n(r)$ via $n_i (r_i) \, r_i^2 = n(r) \, r^2$. Thus, to leading order in the mass-splitting, the mass density profiles are similarly related $\rho_i (r_i) \, r_i^2 \simeq \rho(r) \, r^2$. The final form of the mass density profile is then determined by
\be
\label{eq:rhoupscatter}
\rho(r) \simeq \bigg( \frac{r_i (r)}{r}\bigg)^2 \, \rho_i \big[r_i(r) \big]
~,
\ee
where the function $r_i(r)$ is determined from \Eq{eq:drdt}, as described above. We have found that \Eq{eq:rhoupscatter} accurately captures the salient features found in the ``endothermic up-scattering'' simulation of Ref.~\cite{ONeil:2022szc}. However, note that the above analysis does not incorporate the full set of dynamics, as it only allows for $\x_1 \to \x_2$ conversion and not the reverse process. 

\twocolumngrid
\bibliography{main}

\end{document}